\documentclass[preprint,12pt,authoryear]{elsarticle}

\usepackage{amssymb}
\usepackage{amsmath}

\usepackage{geometry} 
\usepackage{hyperref}
\hypersetup{
    colorlinks=true,
    linkcolor=blue,
    filecolor=magenta,      
    urlcolor=cyan,
}
\usepackage{subfig} 

\journal{arXiv}

\newcommand{\var}{\text{var}}
\DeclareMathOperator{\E}{\mathbb{E}}

\DeclareMathOperator{\N}{\mathbb{N}}
\DeclareMathOperator{\K}{\mathcal{K}}

\DeclareMathOperator*{\argmin}{arg\,min}

\newtheorem{theorem}{Theorem}[section]
\newtheorem{proposition}[theorem]{Proposition}

\begin{document}

\begin{frontmatter}



\title{Consistent second-order discrete kernel smoothing using dispersed Conway-Maxwell-Poisson kernels}


\author{Alan Huang$^1$, Lucas Sippel$^1$, \& Thomas Fung$^2$}

\address{$^1$School of Mathematics and Physics, University of Queensland, QLD, Australia \\ $^2$Department of Mathematics and Statistics, Macquarie University, NSW, Australia}

\begin{abstract}
The histogram estimator of a discrete probability mass function often exhibits undesirable properties related to zero probability estimation both within the observed range of counts and outside into the tails of the distribution.  To circumvent this, we formulate a novel second-order discrete kernel smoother based on the recently developed mean-parametrized Conway--Maxwell--Poisson distribution which allows for both over- and under-dispersion. Two automated bandwidth selection approaches, one based on a simple minimization of the Kullback--Leibler divergence and another based on a more computationally demanding cross-validation criterion, are introduced. Both methods exhibit excellent small- and large-sample performance. Computational results on simulated datasets from a range of target distributions illustrate the flexibility and accuracy of the proposed method compared to existing smoothed and unsmoothed estimators. The method is applied to the modelling of somite counts in earthworms, and the number of development days of insect pests on the Hura tree.

\end{abstract}




\end{frontmatter}


\section{Introduction}
\label{}
Kernel smoothing for continuous probability densities has been covered extensively in the literature \citep[see, for example,][among others]{Rosenblatt1956,Parzen1962,
Deheuvels1977,
Silverman1986,BGK2010}. In contrast, kernel smoothers for discrete distributions have not been explored much, perhaps because the histogram is already a consistent estimator of the underlying probability mass function and exhibits certain optimality properties across samples \citep{KK2011,Kiesse2017}.

\begin{figure}
\centering
\includegraphics[width = 0.48\textwidth, trim = 4 0 5 5, clip]{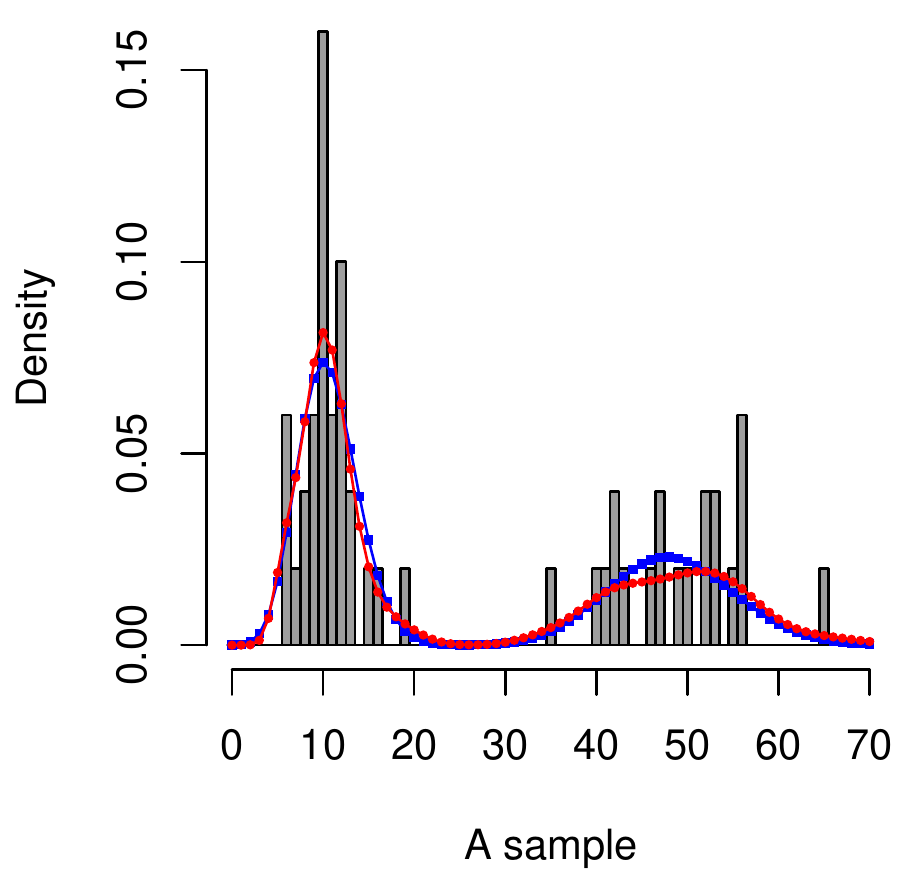}
\caption{Histogram of a sample of $n=50$ counts from a bimodal Poisson mixture distribution, with the Conway-Maxwell-Poisson associated kernel smoothed (red) and true (blue) pmfs overlayed.}
\label{fig:1}
\end{figure}

However, for any given sample of counts, even those with moderately large sample sizes, there are often two undesirable properties of the empirical distribution for estimating the underlying discrete distribution: 
\begin{enumerate}
\item There is a gap within the observed support. For example, in the sample of $n=50$ counts in Figure \ref{fig:1} there were no observations with value 14, 17, 18, or between 20 and 34 inclusively, and so on. To estimate the underlying probability at these non-observed values as exactly zero (i.e., $\hat P(X = x)= 0$ for $x=14,17,18$, and $\hat P(20 \le X \le 34) = 0)$) does not seem reasonable if the underlying distribution is smooth. Therefore, some sort of in-filling may be appropriate for any given sample.
\item  Tail probabilities outside the observed support are necessarily assigned zero probability. For example, in the sample in Figure \ref{fig:1} there were no observations with value below 6 or above 65. To estimate the underlying tail probabilities as exactly zero (i.e., $0 = \hat P(X \le 5) = \hat P (X \ge 66)$) does not seem reasonable if the underlying distribution has no strict upper or lower bound (apart from 0). Thus, some sort of smoothing into the tails may also be helpful for any given sample. 
\end{enumerate}

{\color{black} Additionally, a reviewer pointed out that another important motivating factor for discrete kernel smoothing is that many count datasets arise from discretizing an underlying continuous random variable. For example, in Section \ref{sec:5.2} we consider the developmental period of the spirally whitefly from an egg to adult stages, which could be measured in hours (or even minutes) but is recorded only as integer days. It is therefore expected that the resulting count distribution is ``smooth" because it is a discretization of some underlying continuous density.}

This note introduces a consistent second-order discrete associated kernel smoother for estimating probability mass functions (pmfs) using on the recently developed mean-parametrized Conway--Maxwell--Poisson (CMP) distribution \citep{Huang2017}. This family of discrete distributions behaves similarly to the continuous normal distribution in that it is characterized by a mean (location) and dispersion (scale) parameter, with the two parameters being functionally independent, making it a suitable candidate for a discrete smoother. In particular, CMP kernels are unique amongst discrete distributions in that they can be arbitrarily underdispersed, which is key to obtaining consistency as the sample size increases. Applying the proposed estimator to the sample from Figure \ref{fig:1}, coupled with the automated bandwidth choice prescribed in Section 3, gives the smoothed pmf illustrated by the red curve which offers a markedly improved estimate of the underlying bimodal discrete distribution (blue curve) over the naive histogram estimator.

A summary of the current literature associated with discrete kernel smoothing is given in Section \ref{sec:2}. In Section \ref{sec:3} we construct a new discrete estimator and outline an automated bandwidth selection method based on either minimizing the Kullback--Leibler divergence or via cross-validation. In Section \ref{sec:4} we compare the proposed estimator with existing first- and second-order discrete kernel smoothers of \cite{KK2011} and \cite{Kiesse2017}. In particular, it is demonstrated via extensive simulations that the proposed estimator offers improved small, moderate, and large sample performance compared with existing estimators. The proposed method is then applied to two datasets in Section \ref{sec:5}, the first on the counts of somites in earthworms and the second counting the number of development days of the spirally whitefly on Hura fruit trees.

\section{Discrete Associated Kernels}
\label{sec:2}
The concept of a discrete associated kernel (dak) was introduced in \cite{KK2011} as a general method for estimating discrete pmfs, although earlier ad-hoc methods were described in \cite{MM1999}, \cite{KKZ2007}, and \cite{KZ2010}. This section summarises the current literature on the class of pmf estimators based on daks.
\subsection{Second-order discrete associated kernel estimators}
{\color{black} A second-order dak is a family of pmfs $\left\{K_{xh}(\cdot);\  x \in S_x,\ h \ge 0 \right\}$ on support  $S_x\subseteq\mathbb{N}$ satisfying
\begin{equation}
    \lim_{h \to 0} \E(\K_{xh})= x \quad \text{and} \quad
    \lim_{h \to 0} \var(\K_{xh})= 0\label{dak3}
\end{equation}
for every $x \in S_x$, where $\K_{xh}$ is a random variable with pmf $K_{xh}(\cdot)$ and $h \ge 0$ is a bandwidth parameter.} A second-order dak estimator based on samples $X_1,\ldots,X_n \stackrel{\rm iid}{\sim} f$ is then defined in \cite{KK2011} as
\begin{equation}
    \hat{f}(x) = \frac{1}{n}\sum_{i=1}^n K_{xh}(X_i)\ . \label{diskde}
\end{equation}
Note that unlike continuous kernel density estimators, each kernel here is centred at the point of probability estimation $x$ rather than at each observation $X_i$. For the latter approach, an alternative dak estimator of $f(x)$ is 
\begin{equation}
    \hat{f}(x) = \frac{1}{n}\sum_{i=1}^n K_{X_i h}(x),\label{diskde2}
\end{equation}
where each kernel is centred at the observed values $X_i$ and evaluated at the point of estimation $x$. \cite{KK2011} note that the first formulation is often easier to work with mathematically. However, the second formulation is often computationally  quicker because we need only evaluate each kernel at each unique observation. Note that if the kernel function is symmetric then both formulations are equivalent.

Second-order daks are preferred because they exhibit consistency. Suppose (\ref{dak3}) holds.  Then the main results from \cite{KK2011} give
\begin{equation}
    \lim_{n\xrightarrow{}\infty}\E\{\hat{f}(x)-f(x)\}^2=0 \mbox{ and  }
    \hat{f}(x)\xrightarrow{a.s.}f(x) \text{ for every } x\in \N \ ,
\end{equation}
for any sequence $h \to 0$ as $n \to \infty$. In particular, the usual condition $\lim_{n\xrightarrow{}\infty}\{h+(nh)^{-1}\}=0$ for continuous kernel density estimation is not required for these consistency results.

Although condition (\ref{dak3}) looks like a simple requirement, it is not at all trivial to satisfy for discrete distirbutions. Indeed, only two examples have been presented in the literature, with this paper providing a third and, arguably, only non-trivial family of discrete kernel functions satisfying this requirement.  The first example in the literature is the naive histogram estimator, which is a dak estimator with Dirac-delta kernel functions satisfying (\ref{dak3}) trivially. 
The only other second-order dak estimator that has been proposed in the literature uses the symmetric triangular kernel introduced in \cite{KKZ2007}:
\begin{equation}
    K_{xh}(y) \propto (a+1)^h-|y-x|^h \ , \text{ for } y\in S_x,\label{tke}
\end{equation}
where $h \ge 0$ is the bandwidth, $a$ is a fixed range parameter, and the support of each kernel is the finite set $S_x = \{x, x\pm1,\ldots,x\pm a\}$. Although the triangular dak estimator is consistent, it can offer poor finite-sample performance due to a boundary bias to the left of $\mathbb{N}$ because the kernels can assign probability to negative integers when centred at $0\leq x < a$.  \cite{KZ2010} remedy this by using a modified asymmetric triangular kernel near 0 and the usual symmetric kernel away from 0. However, because each triangular kernel has a finite support either side of its center, it necessarily assigns zero probability to values that are beyond $\pm a$ outside the observed range, or if there is a gap in the observed support that spans more than $2a$. Thus, the undesirables qualities of the naive histogram estimator are still present for triangular dak estimators. The second-order dak estimator based on the CMP distribution in this note does not exhibit such drawbacks.

\subsection{First-order discrete associated kernels}

An alternative to second-order daks is first-order daks, such as the Poisson, binomial and negative-binomial kernels discussed in \cite{KK2011}, which relax the variance criterion in (\ref{dak3}) to 
the less strict condition,
\begin{equation}
    \lim_{h\xrightarrow{}0}\var{(\K_{xh})}\in\mathcal{V}_0,\label{sdk1}
\end{equation}
where $\mathcal{V}_0$ is some neighbourhood of 0 not depending on $x$. While some first-order dak estimators have been demonstrated (via simulations) to perform better than the second-order histogram and triangular estimators for small to medium sample sizes \citep{KK2011}, they lose consistency because they do not satisfy (\ref{dak3}); see \cite{Kiesse2017} for further explanation. 

Of these, the best performing first-order dak is the binomial, and so we will compare our proposed estimator to the binomial dak estimator via simulations in Section \ref{sec:4}. In short, the proposed estimator exhibits improved finite-sample performance over all competing first-order and second-order daks while still retaining consistency for increasingly large samples, thereby offering the best of both worlds.

\section{Conway--Maxwell--Poisson discrete associated kernel smoother}
\label{sec:3}
A recent advancement by \cite{Huang2017} showed that the Conway--Maxwell--Poisson (CMP) distribution, which is a generalization of the Poisson distribution, can be characterized by its mean $\mu \ge 0$ and dispersion $\nu \ge 0$, with the two parameters being functionally independent. The subsequent mean-parametrized CMP distribution has pmf given by
$$
C(x ;\mu, \nu) = \frac{1}{Z(\lambda(\mu, \nu), \nu)}\frac{\lambda(\mu, \nu)^x}{(x!)^\nu} \ , \quad \text{ for } x \in \mathbb{N} \ ,
$$
where $Z(\lambda, \nu) = \sum_{x=0}^\infty \lambda^x/(x!)^\nu$ is the normalizing constant of the distribution, and $\lambda(\mu, \nu)$ satisfies $0 = \sum_{x=0}^n (x-\mu) \lambda^x/(x!)^\nu$. It can be shown that $\nu <1$, $\nu = 1$ and $\nu > 1$ correspond respectively to overdispersion, equidispersion and underdispersion relative to the Poisson$(\mu)$ distribution \citep[see][Section 2]{Huang2017}. Moreover, using the same arguments as in \cite{Huang2020}, the family of mean-parametrized CMP distributions can be shown to be second-order daks for increasingly large $\nu$. Indeed, the CMP distribution is currently the only known generalization of the Poisson that is a second-order dak, making it uniquely suitable for consistent discrete kernel smoothing.
\begin{proposition}
The pmfs $\{C(\cdot; \mu, \nu)\}$ are second-order daks satisfying
$$
    \lim_{\nu \to \infty} \E(C(\cdot; \mu, \nu))= \mu \quad \text{and} \quad
    \lim_{\nu \to \infty} \var(C(\cdot; \mu, \nu))= 0 \ , \text{ for every } \mu \in \mathbb{N} \ .
$$
\end{proposition}

Given samples $X_1,\ldots,X_n$ from some discrete distribution $f$, a CMP dak smoother with bandwidth $h \ge 0$ can then be constructed by placing kernel functions $C(x; \mu, \nu)$ with mean $\mu$ centered at each observation $X_i$ and dispersion index $\nu=1/h$ so that $h \to 0$ implies $\nu \to \infty$:
\begin{equation}
   \hat{f}_{\rm cmp}(x) = \frac{1}{n}\sum_{i=1}^n C(x; X_i, 1/h), \text{ for } x\in\mathbb{N} \ .
   \label{eq:compak}
\end{equation}
By adapting the main results from \cite{KK2011} to the CMP dak estimator (\ref{eq:compak}), we can show that it is consistent for the underlying pmf $f(x)$:
\begin{proposition} Let $h$ be any sequence of bandwidths such that $h \to 0$ as $n \to \infty$. Then,
$$
    \lim_{n\xrightarrow{}\infty}\E\{\hat{f}_{\rm cmp}(x)-f(x)\}^2=0 \mbox{ and  }
    \hat{f}_{\rm cmp}(x)\xrightarrow{a.s.}f(x) \text{ for every } x\in \N \ .
$$

\end{proposition}

\subsection{Bandwidth selection via minimising Kullback--Leibler divergence}
As with any kernel smoothing method, bandwidth selection plays a crucial role in determining finite-sample performance. Here, we consider a simple and computationally efficient bandwidth selection method that is inspired by a discrete analogue to Silverman's rule-of-thumb: optimize for a nominal target discrete distribution but adapt well (or ``fail gracefully") for departures from the nominal distribution.

For a pure count process, the default target would be the Poisson distribution. However, the problem of zero probability estimation both within and outside the observed support is more prevalent for overdispersed distributions, making the Negative-Binomial distribution a reasonable candidate also. To this end, write $f_{\rm pois}(x;\lambda)$ for the pmf of Poisson$(\lambda)$ distribution and $f_{\rm nb}(x; \mu, r)$ for the pmf of the Negative-Binomial$(\mu, r)$ distribution with mean $\mu$ and variance $\mu + \mu^2/r$. Then given a sample $X_1,...,X_n$ from some distribution $f$, the best fitting Poisson distribution would be $\hat f_{\rm pois}$ with $\lambda = \bar{X}$ and for the Negative-Binomial distribution $\hat f_{\rm nb}$ with $\mu = \bar{X}$ and $r = \bar{X}^2/(S^2-\bar{X})$ by the method-of-moments. We then choose the bandwidth $h_{\rm KL}$ that minimizes the worst-case Kullback--Leibler divergence from $\hat f_{\rm pois}$ or $\hat f_{\rm nb}$ to the CMP dak estimator,
\begin{eqnarray*}
h_{\rm KL} = \argmin_{h \ge 0} \left[ \max \left\{{\rm KL} \left(\hat f_{\rm cmp} \| \hat f_{\rm pois} \right), \  {\rm KL} \left(\hat f_{\rm cmp} \| \hat f_{\rm nb} \right) \right\} \right] \ ,
\end{eqnarray*}
where ${\rm KL}(P || Q) = \sum_x P(x) \log \left[P(x)/Q(x) \right]$ for discrete distributions $P$ and $Q$.
We demonstrate in the next section that this simple ``minimax"-type bandwidth selection approach leads to CMP dak estimators that can perform better than the binomial and triangular dak estimators of \cite{KKZ2007} and \cite{KK2011} in terms of the integrated squared error (ISE) for different types of target distributions across small and large sample sizes. This is somewhat surprising, given that the two competing methods employ cross-validated bandwidths specially designed to minimise an ISE-type criterion.

\subsection{Bandwidth selection via cross-validation of predictive probabilities}
Alternatively, when the sample of counts exhibits particularly interesting features such as severe underdispersion or strong left-skewness, a more computationally intensive bandwidth selector via leave-one-out cross-validation can be constructed as follows: for each observation $j=1,2,\ldots,n,$ construct the CMP dak estimator omitting observation $j$, i.e.,
$$
\hat f_{\rm cmp}^{(-j)}(x)  = \frac{1}{n-1} \sum_{i \ne j} C(x, X_i, 1/h) \ ,
$$
and then evaluate the estimated pmf at the omitted observation, $\hat f_{\rm cmp}^{(-j)}(X_j)$. The optimal cross-validated bandwidth $h_{\rm CV}$ is then defined as the value of $h$ that maximizes the likelihood $\prod_{j=1}^n \hat f_{\rm cmp}^{(-j)}(X_j) $, or equivalently, the log-likelihood $\sum_{j=1}^n \log \left(\hat f_{\rm cmp}^{(-j)}(X_j)\right)$. Note that the cross-validation criterion here is based on predictive probabilities of each observation and is distinct from $L_2$ cross-validation that is common in kernel estimation literature. Predictive probabilities are arguably more appropriate for discrete distributions than $L_2$ norms.

As demonstrated next, the cross-validated bandwidth $h_{\rm CV}$ can perform as well as or even better than the minimum Kullback--Leibler bandwidth $h_{\rm KL}$ (at the cost of computational speed).

\section{Simulation study}
\label{sec:4}
\begin{figure}
\centering
\includegraphics[width = 0.95 \textwidth, trim = 3 10 8 15, clip]{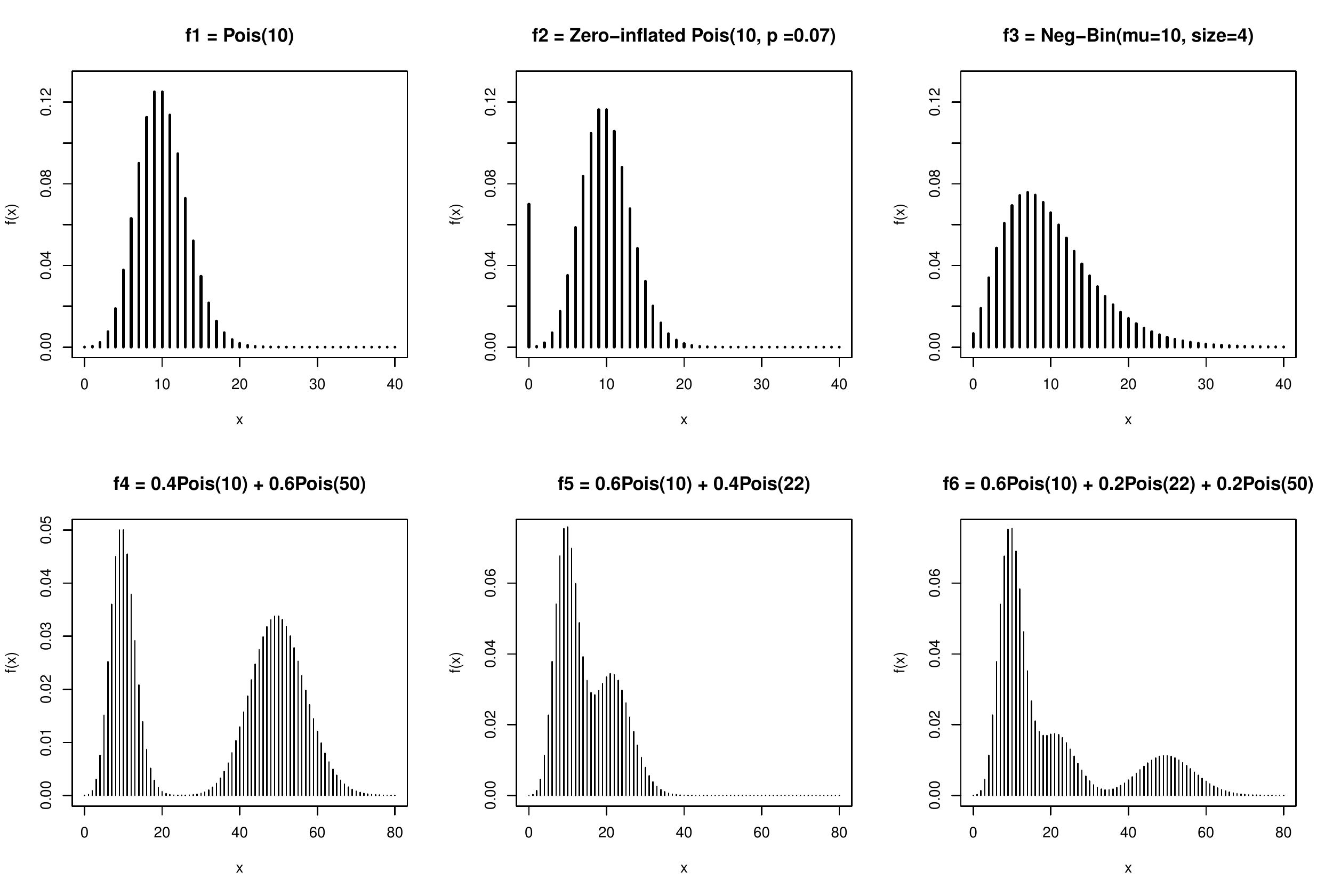}
\caption{Six target distributions used in simulations for comparing accuracy of competing estimators}
\label{fig:target}
\end{figure}

Six target distributions displayed in Figure \ref{fig:target}, covering a range of unimodal, bimodal, trimodal, zero-inflated, equidispersed and overdispersed discrete distributions, were used to compare the accuracy of the CMP dak estimator (\ref{eq:compak}) against the histogram, first-order binomial and second-order triangular dak estimators. To fit the CMP dak estimator we use the \texttt{compak R} package from the authors' Github, which is built on \cite{FAWH2019}'s \texttt{mpcmp} package for fitting mean-parametrized CMP models, while the triangular and binomial daks were fit using the \texttt{Ake R} package by \cite{WSK2015}. The histogram estimator was computed using \texttt{hist} in base \texttt{R}. All computations were carried out on a Windows laptop with an Intel i7-5600 CPU at 2.6GHz and 16GB of RAM.

For each scenario, a sample of $n=20, 50$ or $100$ observations were generated and all four estimators were fit to the same dataset. Following \cite{KK2011}, goodness-of-fit is measured using the ISE 
between the estimated $\hat f$ and the true $f$ pmfs:
$$
{\rm ISE} = \sum_{x=0}^\infty (\hat f(x) - f(x))^2 \ .
$$
This was repeated over $N=1000$ simulations in each setting, and the average ISE 
values, along with their standard deviations, are given in Table \ref{tab:ISE}.

\begin{table}
\centering
\footnotesize
\begin{tabular}{cr|rrrrrrrrrr} 
 & & \multicolumn{10}{c}{estimator} \\
\cline{4-11}
&  & \multicolumn{2}{c}{\underline{histogram}} & \multicolumn{2}{c}{\underline{binomial}}  & \multicolumn{2}{c}{\underline{triangular}} & \multicolumn{2}{c}{\underline{CMP ($h_{\rm KL}$)}} & \multicolumn{2}{c}{\underline{CMP ($h_{\rm CV}$)}} \\ 
Target  & $n$ & mean & sd & mean & sd & mean & sd & mean & sd  & mean & sd \\ 
\hline
$f_1$ &  20 & 46.2 & 19.6 & 14.2 & 9.3 & 15.5 & 11.3 & \textbf{5.3} & 4.7 &  8.6 & 7.9\\
      &  50 & 18.2 &  7.8 &  5.9 & 4.0 &  6.1 &  3.9 & \textbf{2.8} & 2.2 &  3.7 & 3.4\\
      & 100 &  9.3 &  4.3 &  2.5 & 1.5 &  3.2 &  2.4 & \textbf{1.8} & 1.3 &  2.3 & 1.7\\ 
$f_2$ &  20 & 45.8 & 18.9 & 16.4 & 9.6 & 17.8 & 10.8 & \textbf{8.3} & 7.0 & 10.8 & 8.7\\
      &  50 & 18.3 &  7.7 &  6.4 & 4.0 &  8.6 &  4.3 & \textbf{4.2} & 3.0 &  5.2 & 3.7\\
      & 100 &  9.1 &  3.8 &  3.1 & 1.5 &  5.1 &  2.0 & \textbf{2.3} & 1.7 &  2.8 & 2.0\\
$f_3$ & 20 &  47.1 & 14.8 & 14.4 & 6.7 & 15.7 & 8.0 & 8.5 & 4.4 & \textbf{5.7} & 6.2 \\
      & 50 &  18.6 & 5.9 & 5.9 & 3.1 & 6.2 & 3.3 & 4.7 & 2.2 & \textbf{2.7} & 2.2 \\
      & 100 & 9.6 & 3.1 & 2.9 & 1.4 & 3.2 & 1.7 & 3.0 & 1.3 & \textbf{2.6} & 1.2 \\
$f_4$ & 20 & 49.2 & 12.7 & 15.2 & 5.7 & 16.6 & 7.2 & \textbf{4.4} & 2.7 & 5.0 & 3.3 \\
      & 50 & 21.6 & 5.1 & 6.9 & 2.4 & 7.3 & 2.9 & \textbf{2.0} & 1.2 & 2.4 & 1.4 \\
      & 100 & 9.9 & 2.3 & 3.1 & 1.1 & 3.3 & 1.4 & \textbf{1.2} & 0.7 & 1.5 & 0.9 \\
$f_5$ & 20 & 47.6 & 14.8 & 14.7 & 6.2 & 16.4 & 8.3 & 6.2 & 3.6 & \textbf{5.7} & 4.5  \\
& 50 & 18.9 & 5.0 & 5.7 & 2.6 & 6.3 & 2.6 & 3.4 & 1.8 & \textbf{3.0} & 2.1 \\
& 100 &  10.0 & 3.1 & 3.3 & 1.6 & 3.5 & 1.8 & 2.1 & 1.1 & \textbf{1.7} & 1.1\\
$f_6$ & 20 & 46.9 & 12.8 & 14.9 & 5.3 & 15.3 & 6.6 & 5.4 & 3.2 & \textbf{5.0} & 3.2 \\
& 50 &  19.8 & 5.1 & 6.3 & 2.1 & 6.8 & 2.7 & 2.8 & 1.5 & \textbf{2.8} & 1.7 \\
& 100 &  9.9 & 2.7 & 3.1 & 1.1 & 3.4 & 1.4 & 1.8 & 0.9 & \textbf{1.7} & 0.9
\end{tabular}
\caption{Mean and standard deviation of the integrated squared error ($\times 10^{-3}$) between the estimated and true pmfs across $N=1000$ simulations for each setting. The lowest ISE in each setting is displayed in \textbf{bold}.}
\label{tab:ISE}
\end{table}

We see from Table \ref{tab:ISE} that for each target distribution and for any sample size, the CMP dak estimator with either the minimum KL or cross-validated bandwidth always had the smallest ISE. The difference between the two bandwidth selection methods is also minimal in most  cases, and both provide a marked improvement over the naive histogram as well as the triangular and binomial estimators. Additional simulations carried out by the authors for larger samples also demonstrate the consistency of the proposed estimators for increasing sample sizes. 

The average time taken to fit the CMP dak on $n=100$ samples from target distribution $f_6$ was 0.06 seconds when using the KL bandwidth and 3.9 seconds when using cross-validation. Both of these are an order of magnitude faster than the binomial and triangular daks from the \texttt{Ake} package which took on average 17.7 and 22.7 seconds, respectively.

\subsection{Tail probability estimation}

We also examine each competing estimator's ability to smooth into the tails of the distribution. For each target distribution, consider estimating $P(X > x^{(0.99)})$, the probability of observing a value exceeding the top percentile $x^{(0.99)}$, via
$$
\hat P(X > x^{(0.99)}) = 1 - \sum_{x=0}^{x^{(0.99)}}\hat{f} (x) \ ,
$$
where $\hat f$ is a pmf estimator. For samples of size $n=100$ we expect to observe only one realization within this range, and for smaller sample sizes we would expect fewer than one. Hence, $\hat P(X > x^{(0.99)}) = 0$ typically for the histogram estimator and so some sort of smoothing is required for such tail probability estimation. 

The accuracy of tail probability estimation is assessed via the relative error measure,
$$
    r = \bigg{|}\log_{10}\frac{\hat P(X >  x^{(0.99)})}{ P(X >  x^{(0.99)}) }\bigg{|} \ ,
$$
so that $r=0$ when $\hat P(X >  x^{(0.99)}) = P(X >  x^{(0.99)})$ and $r >0$ otherwise. This measure is symmetric in quantifying the magnitude of over or under estimation of the tail probability; for example, estimating either 0.1 or 0.001 for a true tail probability of 0.01 will both be $r=1$ order of magnitude off.

Table \ref{tab:ARE} gives the simulation averages and standard deviations of the relative errors for the binomial, triangular and CMP dak estimators for the six target distributions. The column ``$\infty$" shows the proportion of estimates for which the relative error diverged, and is included for the binomial and triangular estimators because they, much like the histogram estimator, have finite support that often do not contain the upper percentile of the underlying target distribution.

We see from Table \ref{tab:ARE}  that the CMP dak estimator consistently outperforms the binomial and triangular estimators for tail probability estimation, especially after taking into account the proportion of divergent relative errors. In particular, the CMP dak offers significantly improved tail estimation for the bimodal and trimodal target distributions ($f_4,f_5$, and $f_6$) for which the upper percentile can be quite separated from the bulk of the observed data, making tail probabilities particularly difficult to estimate. Analogous simulations for estimating smaller tail probabilities of 0.5\% and 0.1\% were carried out by the authors, with the results further demonstrating the improved accuracy of the CMP dak estimator for smoothing into the tails of the target distribution.

\begin{table}
\footnotesize
\begin{tabular}{cr|ccccccccccc} 
 & & \multicolumn{10}{c}{estimator} \\
\cline{4-11}
&  & \multicolumn{3}{c}{\underline{\quad binomial \quad}}  & \multicolumn{3}{c}{\underline{triangular}} & \multicolumn{2}{c}{\underline{\quad CMP ($h_{\rm KL}$)\quad}} & \multicolumn{2}{c}{\underline{\quad CMP ($h_{\rm CV}$) \quad}} \\ 
Target  & $n$ & mean & sd &  $\infty$ & mean & sd & $\infty$ & mean & sd & mean & sd    \\ 
\hline
$f_1$ & 20 & 1.167 & 0.897 & 11.4\% & 1.278 & 2.964 & 75.2\% & \textbf{0.383} & 0.277 & 0.805 & 0.926 \\
& 50 & 0.670 & 0.563 & 5.9\% & 0.901 & 2.700 & 47.1\% & \textbf{0.325} & 0.214 & 0.585 & 0.673 \\
& 100 & 0.403 & 0.348 & 0.1\% & 0.371 & 1.225 & 25.6\% & \textbf{0.242} & 0.156 & 0.360 & 0.375 \\ 
$f_2$ & 20 & 1.414 & 1.217 & 8.0\% & 2.790 & 4.900 & 70.3\% &  \textbf{0.420} & 0.292 & 0.787 & 0.771 \\
& 50 & 0.744 & 0.761 & 3.5\% & 0.893 & 2.657 & 50.5\% &  \textbf{0.322} & 0.226 &0.554 & 0.572 \\
& 100 & 0.441 & 0.380 & 0.1\% & 0.592 & 1.975 & 26.3\% &  \textbf{0.266} & 0.170 & 0.387 & 0.332 \\
$f_3$ & 20 & 4.292 & 3.476 & 8.2\% & 1.823 & 3.753 & 78.3\% & 1.515 & 0.961 &  \textbf{0.958} & 1.144 \\
& 50 & 2.130 & 2.336 & 3.8\% & 1.363 & 3.531 & 53.6\% & 1.187 & 1.085 &  \textbf{0.614} & 0.788 \\
& 100 & 0.956 & 1.391 & 1.8\% & 0.526 & 1.941 & 31.4\% & 0.773 & 1.020 & \textbf{0.373} & 0.444 \\
$f_4$ & 20 & 2.952 & 2.440 & 1.9\% & 1.298 & 2.914 & 76.2\% & \textbf{0.428} & 0.308 & 0.597 & 0.586 \\
& 50 & 1.480 & 1.600 & 1.6\% & 0.902 & 2.688 & 51.1\% & \textbf{0.346} & 0.231 & 0.442 & 0.383 \\
& 100 & 0.703 & 0.873 & 0.5\% & 0.375 & 1.351 & 28.1\% & \textbf{0.254} & 0.166 & 0.318 & 0.264 \\
$f_5$ & 20 & 2.589 & 2.161 & 3.9\% & 1.813 &3.716 & 78.2\% & 0.824 & 0.671 & \textbf{0.736} & 0.834 \\
& 50 & 1.287 & 1.342 & 2.4\% & 0.982 & 2.861 & 55.5\% & 0.616 & 0.591 & \textbf{0.524} & 0.576 \\
& 100 & 0.691 & 0.784 & 1.2\% & 0.545 & 1.961 & 34.2\% & 0.445 & 0.468 & \textbf{0.363} & 0.358 \\
$f_6$ & 20 & 4.367 & 3.724 & 9.4\% & 1.833 & 3.798 & 77.7\% & 0.814 & 0.680 & \textbf{0.712} & 0.858 \\
& 50 & 2.307 & 2.470 & 1.4\% & 0.955 & 2.734 & 57.7\% & 0.590 & 0.574 & \textbf{0.457} & 0.478 \\
& 100 & 0.976 & 1.447 & 0.6\% & 0.381 & 1.390 & 32.1\% & 0.384 & 0.435 & \textbf{0.332} & 0.334 
\end{tabular}
\caption{Mean, standard deviation, and percentage of $\infty$ of the tail probability relative error over $N=1000$ simulations for each setting. The lowest relative error in each setting is displayed in \textbf{bold}.}
\label{tab:ARE}
\end{table}

\section{Applications}
\label{sec:5}
\subsection{Somite counts in earthworms}
\begin{figure}[]
\centering
\includegraphics[width = \textwidth]{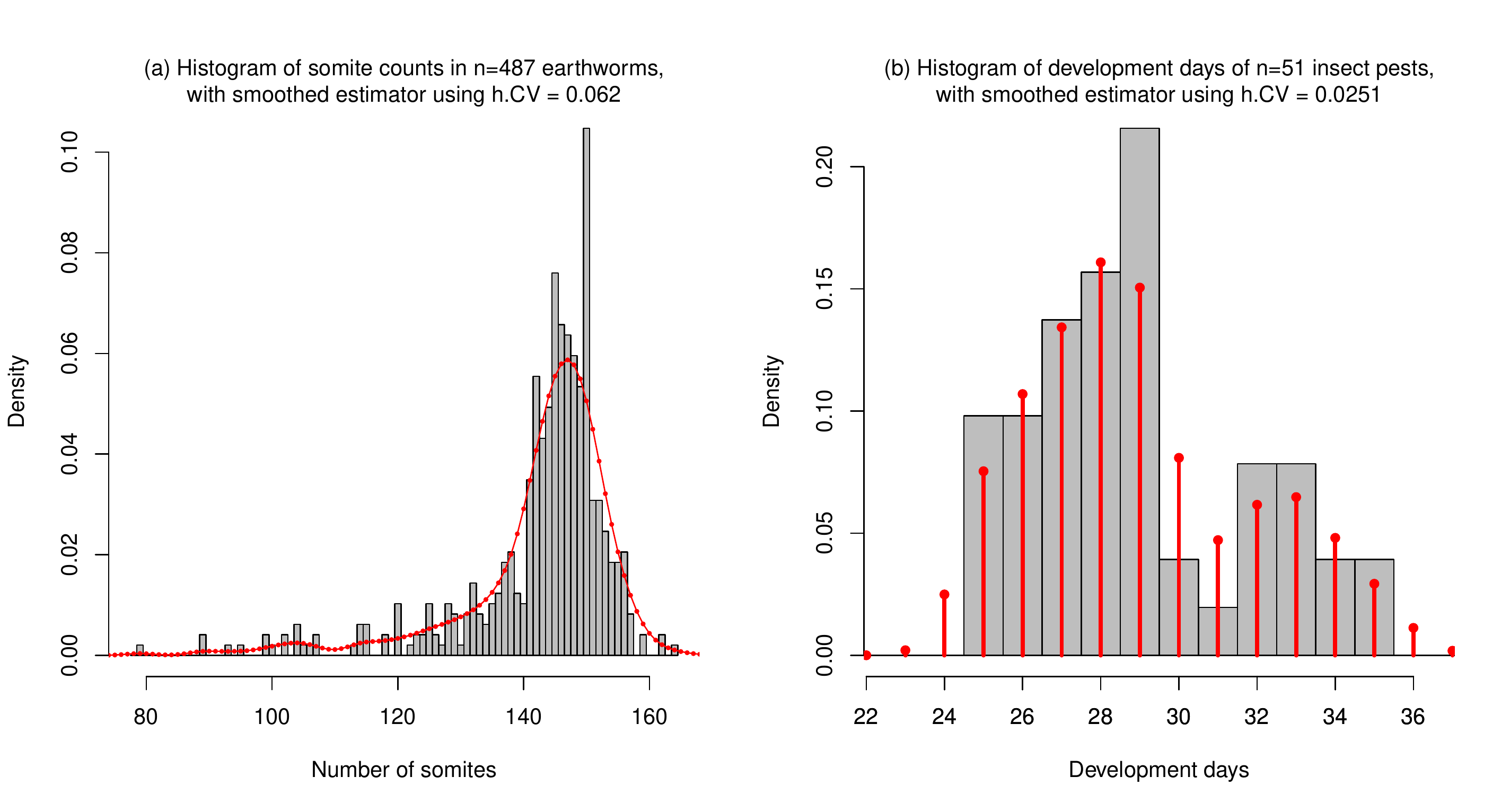}
\caption{ Histograms and smoothed pmf estimators for (a) the number of somites in earthworms; and (b) the count of development days of insect pests on the hura tree.}
\label{fig:somite}
\end{figure}

\cite{Owen2001} describes a dataset from \cite{PF1905} on the number of body segments, known as somites, in common garden earthworms. Figure \ref{fig:somite}(a) shows the histogram of the number of somites on each of 487 worms gathered near Ann Arbor, Michigan, along with the smoothed CMP dak estimate. The strong left-skewness in the data may be due to size-biased sampling, where longer earthworms are easier to observe and gather for counting. The gaps in the counts, between 80--88 or 160--162 say, are surely due to the sampling process and not from any biological or physical reasons, so the proportion of earthworms with somite counts within those ranges should not be exactly zero. To this end, the smoothed CMP dak pmf estimates the proportion of earthworms with somites in these ranges to be $\hat P(80 \le X \le 88) =0.3\%$ and $\hat P(160 \le X \le 162) =0.7\%$. To put these into context, this means that in another sample of $n=487$ earthworms we would expect to see 1.5 earthworms with somite counts between 80--88 and 3.4 with somite counts between 160--162. For tail probabilities, the smoothed CMP dak pmf estimates the proportion of earthworms having fewer than 79 somites to be $\hat P(X \le 78) = 0.1\%$ and more than 164 to be $\hat P(X \ge 165) = 0.2\%$, which allows for the possibility of observing an earthworm with fewer or more somites than in the current sample. Finally, because the CMP kernels are centred around the observed values \citep[and not at the point of estimation as in][and others]{KK2011}, the  smoothed CMP dak estimator preserves the sample mean of the data.

\subsection{Development days of insect pests on Hura trees}
\label{sec:5.2}
\cite{Kiesse2017} describes a dataset on the count of days required for the insect pest (spirally whitefly, {\it Aleurodicus dispersus} Russel) to develop from egg to adult stages while being hosted on the Hura fruit tree plant ({\it Hura crepitans}). A histogram of $n=51$ counts of development days, along with the CMP dak smoothed pmf estimator, is given in Figure \ref{fig:somite}(b).
We see that the histogram, CMP dak, and the binomial and triangular estimators presented in \cite{Kiesse2017} all pick up possible bimodality in the development days. However, here the CMP dak is unique amongst estimators because it is the only one that fits a non-zero probability outside of the observed range of 25 days to 35 days; explicitly, $\hat P(X \le 24) = 2.7\%$ and $\hat P(X \ge 36) = 1.3\%$.
Again, because the CMP kernels are centred around the observed values (and not the target of estimation) the CMP dak estimator preserves the sample mean of the data, a property that is not true of other competing dak estimators.

\section{Conclusion}
A novel discrete kernel smoother has been formulated using mean-parametrized CMP kernels from \cite{Huang2017}. Both the minimal KL and the cross-validated bandwidth selection procedures yield improved estimators for a variety of target distributions. In particular, the ISE measure over the entire support and the accuracy of tail probability estimation for the CMP dak estimator are both demonstrated to be better than existing smoothed and unsmoothed estimators. Further research into improving the accuracy and computational efficiency of the automatic bandwidth selection for the CMP dak estimator is warranted. In particular, we are keen to extend \cite{BGK2010}'s linear diffusion process bandwidth selector to the discrete case. {\color{black} Finally, the tail behaviour of the CMP dak estimator, and how it relates to the tail index $\alpha$ of heavy-tailed pmfs $P(X > x) \propto x^\alpha$, has also been earmarked for future investigation.} 

\appendix

\section*{Acknowledgements} We thank Prof. Dirk Kroese (UQ) and an anonymous reviewer for helpful comments that improved this paper.





\end{document}